\documentclass[preprint,showpacs,preprintnumbers,amsmath,amssymb,nofootinbib]{revtex4}

\usepackage[pdftex]{graphicx}
\usepackage{bm}

\newcommand{\medbox}[1]{\fbox{%
\rule[-10pt]{0pt}{25pt}$\;\;\displaystyle{#1}\;\;$}%
}

\let\ssection=\section
\renewcommand{\section}{\setcounter{equation}{0}\ssection
}
\usepackage{graphicx}
\usepackage{soul} 

\def\parag{\hfil\break} 
\def\kikezd{\parag\underbar}

\newcommand\half{{\scriptstyle{\frac{1}{2}}}}
\def\vF{{\bm{F}}}

\def\2{{\half}}
\def\smallcirc{{\,\raise 0.5pt \hbox{$\scriptstyle\circ$}\,}}
\newcommand{\const}{\mathop{\rm const}\nolimits}
\def\parag{\hfil\break} 
\def\kikezd{\parag\underbar}

\def\br{{\bm{r}}}

\def\beq{\begin{equation}}
\def\eeq{\end{equation}}
\def\beqa{\begin{eqnarray}}
\def\eeqa{\end{eqnarray}}

\def\barray{\left(\begin{array}}
\def\earray{\end{array}\right)}
\def\barraynb{\begin{array}}
\def\earraynb{\end{array}}
\def\benu{\begin{enumerate}}
\def\eenu{\end{enumerate}}

\usepackage[colorlinks=true, pdfstartview=FitV, linkcolor=blue, citecolor=blue, urlcolor=blue]{hyperref} 

\newcommand{\cyan}{\textcolor{cyan}}

\newcommand{\red}{\textcolor{red}}
\newcommand{\blue}{\textcolor{blue}}

\newcommand{\gb}{\colorbox{green}}

\newenvironment{redtext}{\color{red}}{\ignorespacesafterend}
\newenvironment{bluetext}{\color{blue}}{\ignorespacesafterend}
\newenvironment{greentext}{\color{green}}{\ignorespacesafterend}
\newenvironment{magentatext}{\color{magenta}}{\ignorespacesafterend}

\newcommand{\bmag}{\begin{magentatext}}
\newcommand{\emag}{\end{magentatext}}
\newcommand{\bcyan}{\begin{cyantext}}
\newcommand{\ecyan}{\end{cyantext}}

\newcommand{\bblue}{\begin{bluetext}}
\newcommand{\eblue}{\end{bluetext}}
\newcommand{\bred}{\begin{redtext}}
\newcommand{\ered}{\end{redtext}}
\newcommand{\bgreen}{\begin{greentext}}
\newcommand{\egreen}{\end{greentext}}
\newcommand{\bdgreen}{\begin{dgreentext}}
\newcommand{\edgreen}{\end{dgreentext}}

\def\?{\quad{\gb{\fbox{\texttt{?}}\;}}\quad}

\begin{document}

\preprint{\texttt{arXiv:1404.2265v3}  [physics.class-ph]}

\setlength{\baselineskip}{14pt}

\title{The laws of planetary motion, derived from those of a harmonic oscillator (following Arnold)}

\author{
P. A. Horvathy${}^{1}$\footnote{mailto:horvathy@lmpt.univ-tours.fr}
P.-M. Zhang${}^{2}$\footnote{e-mail: zhangpm5@mail.sysu.edu.cn},
}
 
\affiliation{
${}^1${\it Institut Denis Poisson}, Tours University -- Orl\'eans University, UMR 7013
(France).
\\
${}^{2}$ {\it School of Physics and Astronomy}, 
 \\ Sun Yat-sen University Zhuhai 519082, China
\\
}

\date{\today}

\begin{abstract}
Kepler's laws of planetary motion are deduced from those of a harmonic oscillator following Arnold. Conversely the circular orbits through the Earth's center suggested by Galilei are consistent with an $r^{-5}$ potential (as found before by Newton). Both the Kepler/oscillator correspondance and circular orbits are examples of dual potentials.\\
\vskip3mm\noindent
\textit{To be published in Physics Education (India)}
\end{abstract}

\maketitle

\section{Introduction}\label{Intro}

Kepler's laws of planetary motion  state that
\benu
\item
{K-I}: {\it A planet moves on an ellipse, one of whose foci being occupied by the Sun};

\item
{K-II}:  {\it The  vector drawn from
the Sun to the planet's  position sweeps equal areas
in equal times}

\item
{K-III}:  {\it  The squares of the periods are as the cubes of the major axes of the ellipses.}

\eenu

These laws can be deduced from the inverse-square force law and  Newton's equations of motion; the usual proof requires higher mathematics, though \cite{Goldstein}.
  
Analogous statements can be demonstrated using  elementary tools  for a  harmonic oscillator,

\benu
\item
{O-I}: {\it The trajectory under a harmonic force is an ellipse, whose centre is the origin of the linear force};

\item
{O-II}: {\it the vector drawn form the centre of the ellipse to the position sweeps equal areas in equal times};

\item
{O-III}: {\it The periods are independent of the geometric shape of the trajectories.}

\eenu

The similarities are manifest, but neither the differences can  be overlooked.

This aim of note is intended to derive the Kepler laws from those of the oscillator by elaborating the rather concise indications given by Arnold \cite{Arnold}.

\goodbreak
The relation between Keplerian and harmonic motion is first established for circular trajectories.
The second step is a purely geometric correspondence due to Zhukovsky \cite{Zhukovsky}
between two types of ellipses, namely those which appear for an oscillator and those for planetary motion, respectively.
The third step is to extend this purely geometrical correspondence, to a dynamical one. This can be achieved by a rather subtle re-definition of time,
 viewed as a parameter along the trajectory. 

The question asked by Newton in his \textit{Principia} \cite{Principia}
was remarkably \emph{different} from how planetary motion is studied in our time~: instead of solving the dynamical equations for the inverse-square law as we do now, he asked, conversely: \textit{With what form of a central force are the observed trajectories consistent ?}

Newton has actually found general answer for circular motion \footnote{Newton has proved also  that elliptic trajectories can be  consistent with the inverse-square law.}; however his purely geometric argument is difficult to follow today. In sec.\ref{GalNewton} we solve this problem in a particular case considered by Galilei in his \textit{``Dialogo\dots''} \cite{Dialogo}.

\section{Circular motions}\label{circSec}

Let us consider the Kepler problem in the  complex plane, with the Sun fixed at the origin. The  motion of our  planetary is determined by
\begin{equation}
    \ddot{z}=-fM\frac{z}{|z|^3},
    \label{keplereq}
\end{equation}
where the ``dot'' means derivation w.r.t. newtonian time, $\dot{(\,\cdot\,)}=\displaystyle\frac{\ d}{dt}$\,. Our aim is to recover the usual Keplerian trajectories by solving eqn. (\ref{keplereq}). 

We start our investigations with a (very) special case, namely with motion along a  circle of radius $A=\const$. The latter can be parametrised by the angle $\theta$,
$
    z(t)=A\,e^{i\theta(t)} \,;
$ 
then (\ref{keplereq}) requires
\begin{equation}
    i\ddot{\theta}-\left(\dot{\theta}\right)^2=-\frac{fM}{A^3}\,.
    \label{angleeq}
\end{equation}
From the vanishing of the imaginary part we infer that the motion is uniform along the circle, $\ddot{\theta}=0$,
and from the real part we deduce the angular velocity,
\beq
\dot{\theta}=\sqrt{fM/A^3} \equiv \Omega.
\label{oscifreq}
\eeq

It is important that the trajectory studied here can also be viewed  as that of a harmonic oscillator, constrained to move on a circle.
Again using a complex coordinate, $w$, the equation of motion of the planar oscillator are
\begin{equation}
    w''=-\Omega^2w,
\label{oscilleq}
\end{equation}
where  $\Omega$ (assumed real) is the frequency of the oscillator
and the ``prime'' denotes derivation w.r.t. ``oscillator time'', $\big(\,\cdot\,\big)'=\frac{\ d}{d\tau}$.
Then $w(\tau)=B\,e^{i\gamma(\tau)}$ yields 
$i\gamma^{\prime\prime}-({\gamma}^{\prime})^2=-\Omega^2$ which is (\ref{angleeq}),
provided  the angle $\theta$ is identified with $\gamma$, 
$t$ with $\tau$ and  $\Omega^2$ with $fM/A^3$.
 
 Below we extend this correspondence to general motions.

\section{The planar oscillator}\label{oscill}

The motions of a planar oscillator are readily determined. Decomposing (\ref{oscilleq}) into real and imaginary parts, we observe that they perform independent
harmonic motion. With an appropriate choice of the parameters,
\begin{equation}
    w(\tau)=a\cos\Omega\tau+ib\sin\Omega\tau.
    \label{osmot}
\end{equation}
Thus $w(\tau)$ describes an ellipse, whose centre is  the origin and has major and minor axes $a$ and $b$, respectively. 

The areal velocity is half of the conserved angular momentum,
\begin{equation}
    I_{osc}=|w|^2\,\frac{d\gamma}{d\tau}={\rm const}\,.
    \label{tertor}
\end{equation}
The period, 
\begin{equation}
    T_{osc}=\frac{2\pi}{\Omega},
    \label{osciper}
\end{equation}
is independent of the geometrical data of the orbit, as stated.

 In conclusion, we have proved the ``Kepler'' laws O-I--O-II--O-III  of the oscillator-motion.
\goodbreak

We record for further use the expression
for the (conserved) energy in terms of the geometric data of the trajectory,
\begin{equation}
    E_{osc}=\half\left(|w'|^2+\Omega^2|w|^2\right)
    =\half\Omega^2(a^2+b^2) > 0.
    \label{oscenerg}
\end{equation}

\section{Some (complex) geometry \cite{Arnold}}\label{komplex}

Let $u$ denote a complex variable and let us consider 
the so-called ``Zhukovsky-map'' of the complex plane,
\begin{equation}
    u\mapsto w=u+\frac{1}{u}\,.
    \label{elso}
\end{equation}
\goodbreak

\kikezd{Lemma 1}. {\it  The image of a circle of radius $\rho>0$ whose centre is the origin of  the $u$-plane is an \emph{ellipse},  whose centre is the origin of the  $w$-plane. The foci of the ellipse are at the points $\pm2$.}
\vskip3mm

Proof~: if $u=\rho e^{i\phi},\, , 0\leq\phi\leq2\pi$, then
\begin{equation}
w=\left(\rho+\frac{1}{\rho}\right)\cos\phi
+
i\left(\rho-\frac{1}{\rho}\right)\sin\phi,
\label{zsuk}
\end{equation}
which is the equation of an ellipse centered at  $w=0$ and has major and minor axes 
$
a=\rho+\rho^{-1} \geq2
$ and $
b=\rho-\rho^{-1}\geq0,
$
respectively.  
 The foci are at $\sqrt{a^2-b^2}=\pm2$ from the centre. 
The angles $\phi$ and  $\gamma={\rm arg} w$ are related as $\tan \gamma=\frac{\rho^2-1}{\rho^2+1}\tan \phi$.

 When $\rho\to1$  the ellipse degenerates to the
segment $[-2,2]$ on the real axis.
If the radius of the $u$-circle is changed from  $\rho\neq1$ to $\rho^{-1}$ then the image describes, by (\ref{zsuk}), the same ellipse reflected on the real axis.
We can, therefore, chose $\rho>1$, i.e., to restrict the  mapping  $u\mapsto w$ to the exterior of the unit circle.

Let us now consider, following
Levi-Civita  \cite{LC} the mapping of the $w$ plane onto the $z$-plane 
\begin{equation}
    w\mapsto z=w^2.
  \label{LeCi}
\end{equation}
\vskip-5mm
\kikezd{Lemma 2}. {\it The image under (\ref{LeCi}) of the previously constructed $w$-ellipse is an ellipse
in the $z$-plane,  whose left focus is the origin of the $z$-plane. Conversely, each ellipse of the $z$-plane with one focus at the origin is obtained (after a suitable rotation) as the square of an ellipse, whose centre is the origin of the $w$-plane}.

\vskip3mm
This follows from 
$ 
z=w^2=2+u^2+u^{-2}~:   
$ 
putting $u=\rho e^{i\phi}$ yields,  
\begin{equation}
z=2+\left(\rho^2+\frac{1}{\rho^2}\right)\cos2\phi
+
i\left(\rho^2-\frac{1}{\rho^2}\right)\sin2\phi,
\label{zellips}
\end{equation}
cf. (\ref{zsuk}),
which is again an ellipse, whose axes are
$ A=\rho^2+\rho^{-2},
$ and $
B=\rho^2-\rho^{-2},
$ 
and is shifted by $2$ units to the right. Its left focus is at $z=0$.

\begin{figure}
\begin{center}
\includegraphics[scale=.42]{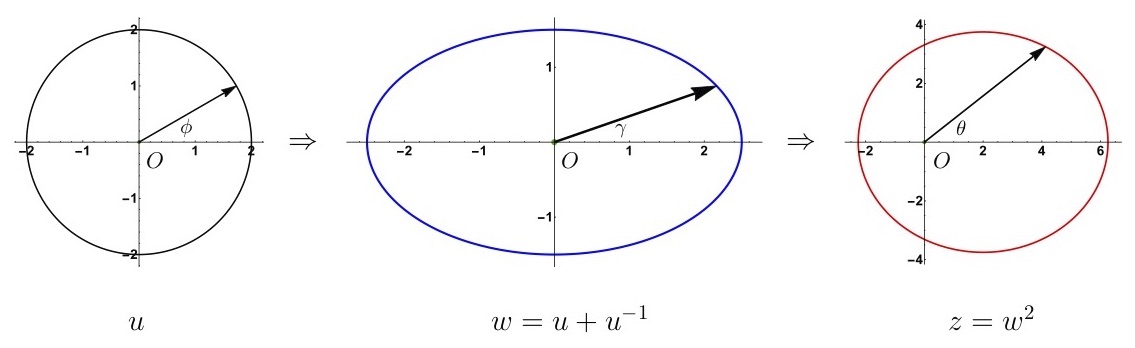}
\vspace{-15mm}
\end{center}
\caption{\it While ${\bf u}$ describes the {\bf circle} of radius $\rho\neq1$ of the complex plane,
$\blue{\bf w=u+u^{-1}}$ moves on an \blue{\bf ellipse} centered at the origin of the $\blue{\bf w}$-plane, and $\red{\bf z=w^2}$ descibes another \red{\bf ellipse} one of whose foci is the origin of the $\red{\bf z}$-plane.
}
\label{Zsukfig}
\end{figure}

While  $u$ goes around the $u$-circle once,  $w$ describes  the ellipse with centre at $w=0$ also once.  $z$ however describes  its ellipse \emph{twice}~: 
 while $w$ describes half of an ellipse, 
$z$ describes the full image-ellipse.   Thus $\phi$ can be restricted to $0<\phi<\pi \Rightarrow 0< \gamma<\pi  $. 
The $z$-perihelion and  aphelion points are, in particular, the images of the end points of the minor and resp. major axes of the $w$-figure, $\phi=\pi/2$ and $\phi=0$, see Fig.\ref{Zsukfig}.
\goodbreak
Let us note that while the Zhukovsky map (\ref{elso}) 
yields ``true'' $w$-ellipses  but no circles ($a > b)$, the  Levi-Civita map (\ref{LeCi}) would work with no such restriction. The image of a $w$-circle  $ae^{i\gamma}$ would be simply the $z$-circle $a^2e^{i2\gamma}$. The Zhukovsky map merely provides a convenient parametrization for the image.

Every $z$-ellipse centered at the origin can be obtained by a suitable rotation and dilation of the one considered here \cite{Arnold}.

\section{The oscillator - Kepler correspondence}\label{oscitransf}

The geometric construction  of the previous section, and eqn. (\ref{LeCi}) in particular,
swaps \emph{oscillator trajectories} with 
\emph{Keplerian} ones. One can wonder if the correspondence can  be extended also to the \emph{dynamics}. At first sight, the answer seems to be negative: differentiating 
$z=w^2$ w.r.t. $\tau$  the eqn. of motion (\ref{oscilleq}) of $w$ yields a rather complicated expression which is manifestly different from (\ref{keplereq}).
This mismatch is also seen by expressing Kepler's areal velocity [i.e. half of the angular momentum] derived using that of the oscillator, 
\beq
I_{K}=|z|^2\theta'=2|z|I_{osc}
\label{IpIo}
\eeq
[where we used $\theta=2\gamma$], 
which \emph{can not be a constant of the motion} therefore, unless  $|z(t)|$ is a constant i.e., with the exception of  circular trajectories.
In other words, Kepler's 2nd law is not in general satisfied.
 
But eqn. (\ref{IpIo}) shows also the way to resolve the contradiction: \emph{time}, or more precisely 
\emph{the parameter used along the trajectories, should be redefined}. Let us indeed consider an arbitrary
oscillator-motion $w(\tau)$, and define the new ``time'' along the trajectory as
\begin{equation}
    t=\int|w(\tau)|^2d\tau
    \quad\Longrightarrow\quad
    \frac{\ d}{dt}=\frac{1}{|w(\tau)|^2}\frac{\ d}{d\tau}
    \label{newtime}
\end{equation}
i.e., $\dot{(\,\cdot\,)}=|w|^{-2}(\,\cdot\,)'$.
Then, using (\ref{oscilleq}),  
\begin{equation}
\ddot{z}=\frac{1}{w\bar{w}} \frac{\ d}{d\tau}
\left(\frac{1}{w\bar{w}}\frac{\ d}{d\tau}w^2\right)
=
-2\left(|w'|^2+\Omega^2|w|^2\right)\frac{1}{w\bar{w}^3}
=
-{4E_{osc}}\frac{w^2}{|w|^6}\,.
\label{eqmottr}
\end{equation}
Here $E_{osc}$ is the oscillator-energy in eqn. (\ref{oscenerg}), which remains constant along the trajectory. Then, putting  $z$ for
$w^2$  yields eqn.  (\ref{keplereq}) of planetary motion with the identification
\begin{equation}
    4E_{osc}=fM. 
    \label{Eosfm}
\end{equation}    

Hence, to each oscillator-trajectory is associated a Keplerian trajectory, whose gravitational force is four times the oscillator-energy, and vice versa. The inverse of (\ref{LeCi}-\ref{newtime})
\begin{equation}
w=\sqrt{z},\quad 
\frac{\ d}{d\tau}=\sqrt{z\bar{z}}\,\frac{\ d}{dt}
\quad\Longrightarrow\quad 
w''=\frac{1}{2}\left(\frac{1}{2}|\dot{z}|^2-\frac{fM}{|z|}\right)w,   
    \label{inversetr}
\end{equation}
which
is the eqn of motion of an oscillator, whose frequency (square) is proportional to the energy of the Keplerian trajectory,
\begin{equation}
    w''=-\Omega^2\,w,\quad
\Omega^2=-\frac{1}{2}E_{K},
\label{ofrek}
\qquad
   E_{K}=\frac{1}{2}|\dot{z}|^2-\frac{fM}{|z|}=
   -\frac{fM}{2A}\,,
\end{equation}
where $A$ denotes the major axis of the Keplerian ellipse.
Let us stress that the transformation is defined
along trajectories only: (\ref{inversetr})  associates a \emph{different} oscillator-trajectory to each Keplerian orbit. For elliptic Kepler motions the frequency $\Omega$ is real, consistently with $E_K<0$.

\section{The Kepler laws}\label{Keplerder}

Now we derive the Kepler laws, KI-II-III, from  those,
OI-II-III, of the oscillator.

Firstly, the Keplerian orbits being images of the oscillator orbits  satisfy KI.

Secondly,
\begin{equation}
    I_{K}=|z|^2\frac{d\theta}{dt}
    =
    |w|^2\frac{d(2\gamma)}{d\tau}=2I_{osc},
    \label{bolyoscter}
\end{equation}
so that the areal velocity of the planet, $\half I_{K}$,
is itself a constant of the motion; this is KII.

Finally, let denote $T_{K}$  and $T_{osc}$ the periods
of the Keplerian resp. oscillator motions.
From the conservation of the areal velocities, 
$
\pi ab=\half I_{osc}T_{osc},
\;
\pi AB=\half I_{K}T_{K},
$
because the area of an ellipse is $\pi$-times the product of its two axes. Then, using 
$T_{osc}=2\pi/\Omega$,
$$
    T_{K}=\pi\left(\frac{AB}{ab}\right)\frac{1}{\Omega}.
$$
As seen in Sect.  \ref{komplex}, 
$a=\rho+\rho^{-1}$, $b=\rho-\rho^{-1}$ and therefore
 $AB/ab=A$.
But
$a^2+b^2=2(\rho^2+\rho^{-2}\big)=2A$. Then from the expression (\ref{oscenerg}) for the energy of an
oscillator-ellipse we infer, using  (\ref{Eosfm}),
that a Keplerian trajectory with 
major axis $A$ has frequency and period
\begin{equation}
\Omega=\sqrt{\frac{E_{osc}}{A}}=\sqrt{\frac{fM}{4A}}\,
\quad\Rightarrow\quad
    T_{K}=2\pi\sqrt{\frac{A^3}{fM}}
    \label{boly}
\end{equation}
which  plainly implies
Kepler's III. law.

\section{Scattered motions}\label{scatt}

As mentioned before, the formula for the (negative) total energy of planetary motion in (\ref{ofrek}) is only valid for bound (elliptic) motions; for  unbound (parabolic or hyperbolic) motions, the energy is zero resp. positive. Our oscillator-planet correspondence can be extended to these  motions, and in fact also to those in Rutherford's experiment with repulsive interaction.

\begin{figure}
\hskip4mm
\includegraphics[scale=.36]{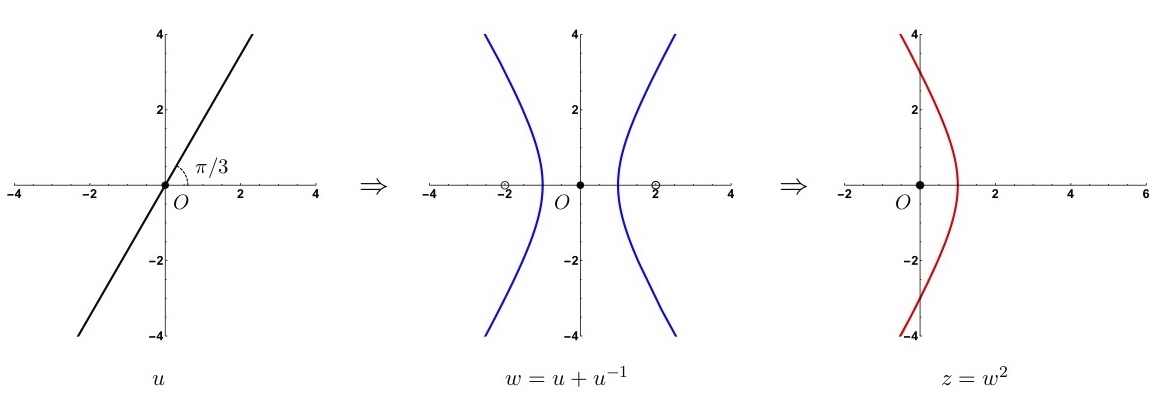}
\vspace{-5mm}
\caption{\it The image of a {\bf straight line}  through the origin of the  $\bf{u}$-plane with inclination $\pi/4 < \phi  < \pi/2$ is a \blue{\bf Zhukovsky hyperbola}, whose centre is the origin of the  $\blue{\bf{w}}$-plane.
Squaring the latter provides us with the \red{\bf left} branch of a hyperbola in the ${\bf{z}}$-plane, whose inner focus is ${z=0}$.
}
\label{lefthypfig}
\end{figure}

\begin{figure}
\hskip-8mm
\includegraphics[scale=.38]{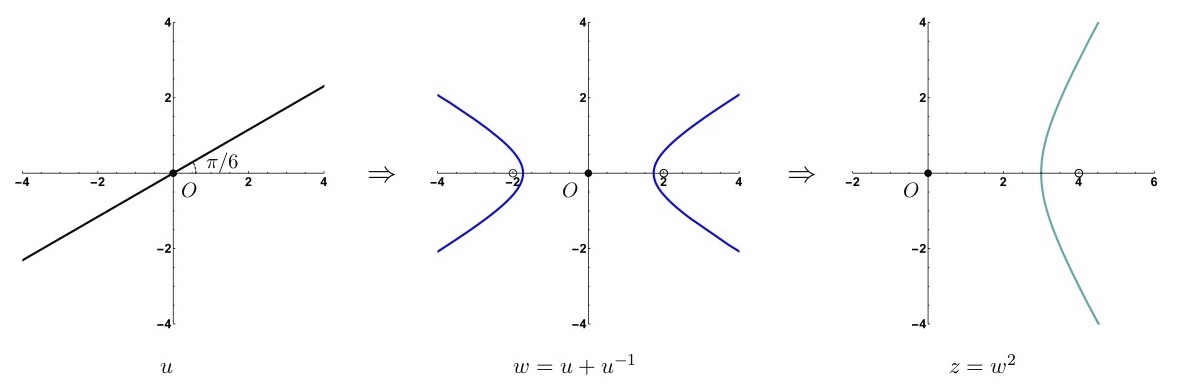}
\vspace{-6mm}
\caption{\it The image of a {\bf straight line} through the origin of the  $\bf{u}$-plane with inclination $0  < \phi  < \pi/4$ is a \blue{\bf Zhukovsky hyperbola}, whose centre is  the origin of the $\blue{\bf{w}}$-plane.
Squaring the latter provides us with the \cyan{\bf right}-branch of a hyperbola in the $z$-plane in the ${\bf{z}}$-plane, whose outer focus is ${z=0}$.
}
\label{righthypfig}
\end{figure}

Let us start with the geometry. We use again   the sequence
$$
u\mapsto w=u+u^{-1} \mapsto z=w^2;
$$
the only difference with the previous case is that the
variable $u$ now describes, instead of a circle as on Fig.\ref{Zsukfig}, a
\emph{straight line through the origin}
, as shown on Figs. \ref{lefthypfig}-\ref{righthypfig}.
For $u=\rho\,e^{i\phi}$ 
\;  $w$ is formally again (\ref{zsuk}),
but now it is $\rho\neq0$ which varies (between $-\infty$ to $\infty$), while the angle is kept fixed, $\phi =\const$. 
Then from
\beq
\left(\frac{\Re(w)}{\cos\phi}\right)^2 -
\left(\frac{\Im(w)}{\sin\phi}\right)^2 
=
\big(\rho+\rho^{-1}\big)^2-\big(\rho-\rho^{-1}\big)^2=4
\label{whyperbola}
\eeq
we infer that (for $\phi \neq 0,\, \pi$) 
we have now a \emph{hyperbola} with real and imaginary axis
\beq 
a=\vert2\cos\phi \vert
\quad\hbox{and}\quad
b=\vert2\sin\phi \vert\,,
\label{hypaxes}
\eeq 
respectively.
The foci are at $\pm2$ because $a^2+b^2=4$.
Fixing $\phi $ between zero and $\pi/2$ and varying $\rho$ from $-\infty \to \infty$, the full Zhukovsky hyperbola is obtained~:
when $0<\phi <\pi/2$, then, for $\rho>0$ we have $\Re(z)\geq2$ and we get the right branch of the $w$-hyperbola. For $\rho<0$ we get instead its left branch, $\Re(z)\leq-2$. If $\phi =0$, $w$ describes two half-lines ($\Re(z)\geq2$ resp. $\Re(z)\leq-2$) of the real axis; for
$\phi =\pi/2$ we get the imaginary axis. 

Choosing $\phi $ between $\pi/2$ and $\pi$ would yield once again the same figure, but in the opposite order. It is therefore enough to choose $\phi $ in the interval  $0\leq \phi \leq\pi/2$. 

The square of the Zhukovsky hyperbola is again (\ref{zellips}) 
which belongs again to a hyperbola shifted by $2$ to the right. This follows from
\beq
\left(\frac{\Re(z)-2}{\cos2\phi}\right)^2 -
\left(\frac{\Im(z)}{\sin2\phi}\right)^2 
=\big(\rho^2+\rho^{-2}\big)^2-\big(\rho^2-\rho^{-2}\big)^2
=4\,.
\label{zhyperbola}
\eeq
This expression is symmetric with respect to changing the sign of $\rho$, therefore we can restrict ourselves to $\rho>0$.
  The real and  imaginary axis  are
$ 
  A=\vert2\cos2\phi \vert
$ {and} $
  B=\vert2\sin2\phi \vert ;
$ 
its {left} focus is the origin.

The subtlety is that choosing the constant parameter $\phi $ between $\pi/4$ and $\pi/2$, we have $\Re(z)\leq2$ so we get only  the \emph{left-branch} of the hyperbolae  cf. Fig.\ref{lefthypfig}; fixing
 $\phi$ instead between zero to $\pi/4$ then
$\Re(z)\geq2$ so that $z$ describes the 
\emph{right-branch}, cf. Fig.\ref{righthypfig}. 

Turning to the dynamics, the oscillator-Kepler correspondance, (\ref{LeCi}-\ref{newtime}) and (\ref{inversetr}) are formally the same as before, except that for positive planet-energy the oscillator-frequency (\ref{ofrek}) is imaginary, $\Omega^2<0$. 
Let us assumme therefore that the linear force is repulsive (also called an inverted oscillator). Then  
\begin{equation}
    w''=+\vert\Omega\vert^2w
    \qquad\Longrightarrow\qquad    w(\tau)=a\cosh\vert\Omega\vert\tau+ib\sinh\vert\Omega\vert\tau
\label{reposceq}
\end{equation}
cf. (\ref{osmot}), which is a Zhukovsky-hyperbola with its centre at the origin.

The potential energy of the repulsive ocillator is negative,
$-\half|\Omega|^2|w|^2$. Therefore the total energy, 
\begin{equation}
    E_{osc}=\half\left(|w'|^2-|\Omega|^2|w|^2\right)
    =\half|\Omega|^2(-a^2+b^2)
    \label{toscen}
\end{equation}
[cf. (\ref{oscenerg})] can  be either positive or negative, depending on whether the imaginary or the real axis is longer.

Our investigations in Sec.\ref{oscitransf} are still valid, and still yield motion with inverse-square force law. When the oscillator-energy is positive, the Kepler problem with $fM=E_{osc}>0$ is obtained, with a hyperbolic trajectory. $E_{osc}>0$ means, moreover, that the imaginary axis is the longer one, $b>a$. 
By (\ref{hypaxes}) this happens  when $\pi/4<\phi <\pi/2$ and then the image 
 is the \emph{left-branch}, -- the one which turns \emph{towards the Sun} as it should for attractive interaction. This is what happens for
most non-periodic comets observed in the solar system. 

What is the use of ``oscillator'' motions with negative energy, $E_{osc}<0$ ? The eqn. of motion in this case, 
\begin{equation}
    \ddot{z}=(-E_{osc})\frac{z}{|z|^3}
    \label{Cmozg}
\end{equation}
cf.  (\ref{eqmottr}),  describes the motion with a repulsive inverse-square 
force, as in the \emph{Rutherford experiment}, when light $\alpha$-particles with charge $q$ are scattered on a heavy atomic nucleus with charge of the same sign $Q$.
The interaction is repulsive, described by (\ref{Cmozg}), with the correspondence
\begin{equation}
    4E_{osc}=-qQ.
    \label{tazon}
\end{equation}

However, by (\ref{toscen}) having  negative  oscillator-energy implies  the real axis being the longer one. But this happens when $0<\phi<\pi/4$ -- i.e., when squaring yields the \emph{right-branch} of the $z$-hyperbola, 
shown on Fig.\ref{righthypfig}. In this case the $\alpha$-particle is \emph{turned away} from the  nucleus, consistently with the repulsive interaction..

Conversely, since the energy is always positive for the repulsive inverse-square force,
\begin{equation}
    E_{Coulomb}=\frac{1}{2}|\dot{z}|^2+\frac{qQ}{|z|}>0
    \label{Cenerg}
\end{equation}
the inverse transformation  (\ref{inversetr}) associates
to the Coulomb-Rutherford problem a repulsive linear system, whose (imaginary) frequency is determined by the Coulomb-energy (\ref{Cenerg}).

Parabolic orbits have vanishing energy;  their associated ``oscillator-frequency" is therefore
$\Omega=0$  by  (\ref{ofrek}) -- i.e., $w$ moves freely along a straight line. If the motion does not
go through the origin, then one can achieve with a suitable dilation and rotation that the trajectory  be 
$w(\tau)=i+\tau$. Then
\begin{equation}
    z=w^2=(\tau^{2}-1)+2\tau i
\end{equation}
is the equation of a ``horizontally lying'' parabola, as seen from $y^2=4x+4$.
The vertex of the parabola is at $x=(-1)$ and its focus 
at the origin $z=0$. If our straight line does go through the origin, its image degenerates into a half-line.

\goodbreak
\section{Galilei and Newton}\label{GalNewton}

\begin{figure} [h]
\includegraphics[scale=0.21]{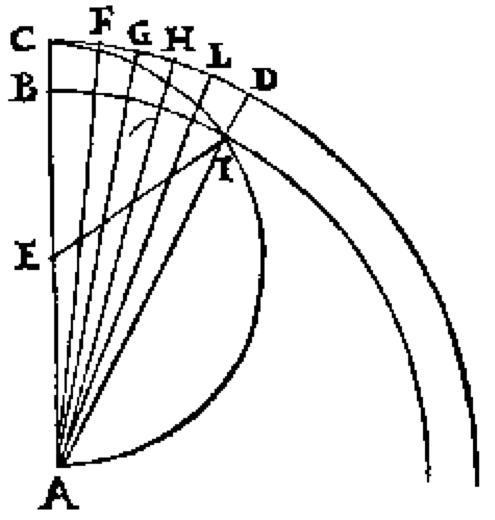}\vskip-5mm
\caption{{\it Figure copied from Galilei's ``Dialogo\dots ''} \cite{Dialogo}~: who suggested that {\it a body dropped from $C$ would follow a circular trajectory which passes through the center $A$ of the earth.} 
\label{DialogoFig}}
\end{figure}

Galileo Galilei, in his ``Dialogo\dots '' \cite{Dialogo}
 suggests that a body dropped from a tower (say the Leaning Tower of Pisa) on  rotating Earth would   follow a circular trajectory which, (if it was not stopped), would pass through the center of the Earth, cf. Fig.\ref{DialogoFig}.\vskip-3mm

What would be the law of gravitation consistent with Galilei's statement~? 

\begin{figure} [h]
\includegraphics[scale=0.35]{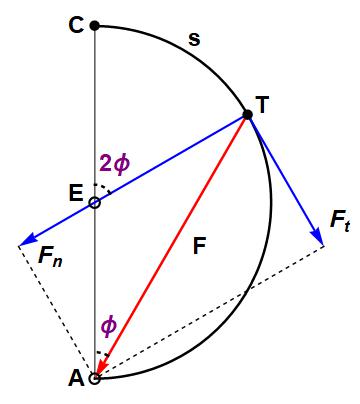}\vskip-6mm
\caption{The force is  directed towards $A$, the center of the Earth. 
\label{Newtoncircle}}
\end{figure}

Consider a body (of unit mass) dropped from  $C$ and assume that it follows a half-circle which passes through  $A$, the Earth' center. Its diameter is thus the radius of the earth, $2\rho=AC=R$, see (Fig.\ref{Newtoncircle}).
The Earth attracts the body by a force $\vF$ 
directed, for symmetry reasons, towards the center, $A$, of the Earth; assume that $F=|\vF|$ is proportional to some power of the distance,
\begin{equation}
F = |\vF|=\gamma\, r^\mu,
 \label{ero}
\end{equation}
where $\gamma$ and $\mu$ are constants to be determined. 
Decompose  $\vF$ into components~: $\vF_n$, the normal component, is  directed towards the center, $E$, of the trajectory and determines the centripetal acceleration, while the tangential component, $F_t$, determines the tangential acceleration,
\begin{equation}
\frac{v^2}{\rho}=F_n,
\qquad
a=F_t,
\end{equation}
where $v$ is the velocity at $T$ and $a=\dot{v}$.
Denoting the angle $CAT$ by $\phi$ and the arc length $CT$ by $s$, we have
$
F_n=F\cos\phi,\, 
F_t=F\sin\phi,\,
s=\rho\,2\phi=R\phi
\,\Rightarrow\,
v=\dot{s}=R\dot{\phi}\,,
a=R\ddot{\phi}.
$ 
Thus we have to solve
\begin{eqnarray}
(\dot{\phi})^2=\frac{F}{2R}\,\cos\phi,
\qquad
\ddot{\phi}=\frac{F}{R}\,\sin\phi\,.
\label{szoggyors}
\end{eqnarray} 
Using (\ref{ero}) and  $r=AT=2\rho\cos\phi=R\cos\phi$ yields,
\begin{eqnarray}
\dot{\phi}=\sqrt{\frac{\gamma}{2}}\,R^{\frac{\mu-1}{2}}
\cos^{\frac{\mu+1}{2}}\phi
\;\Rightarrow\;
\ddot{\phi}=-\left(\frac{\mu+1}{2}\right)
\sqrt{\frac{\gamma}{2}}\,R^{\frac{\mu-1}{2}}
\cos^{\frac{\mu-1}{2}}\sin\phi\cdot\dot{\phi}\,.
\label{szogseb}
\end{eqnarray}
Then comparing with (\ref{szoggyors}) leaves us with an identity, \emph{provided},
\begin{equation}
\medbox{
\mu=-5 \quad\hbox{i.e.}\quad
 F(r) = \gamma\,r^{-5}\,.}
 \label{-5force}
\end{equation}
The body's initial velocity is 
that of the Earth,
$ 
\dot{\phi}(0)=\omega=2\pi/day.
$ 
By (\ref{szogseb}) 
$ 
\gamma=2R^6\omega^2.
$ 

In fact, the problem  can be further generalized. Let as assume that a point particle moves along a known planar trajectory  $r=r(\phi)$ under a central force, which is directed towards  the origin. The force depends on the distance and the azimuthal angle, $\vF=\vF(r,\phi)=F(r,\phi)\hat\br$. What  force law be consistent with the trajectory ?

Angular momentum conservation  implies that the trajectory lies in the plane and that $r^2\dot{\phi}=h=\const$. 
The radial component of the eqn of motion is 
\beq
m(\ddot{r} - r\dot{\phi}^2)=F(r,\phi),
\eeq
implying
\beq
\medbox{
F(r,\phi)=m\left(\ddot{r}-\frac{h^2}{r^3}\right) = -\frac{mh^2}{r^2}\left(\frac{d^2\big(\frac{1}{r}\big)}{d{\phi}^2}+\frac{1}{r}\right)\,.}
\label{forclaw}
\eeq
For the Kepler ellipsis
\beq
r=\frac{p}{1+\epsilon \cos\phi}
\quad\Rightarrow \quad 
F(r,\phi)=-\frac{mh^2}{p}\frac{1}{r^2}\,.
\label{KNforce}
\eeq
For circular motion, taking the  force centrum as the origin,  \eqref{forclaw} implies
\beq
F(r,\phi)=-8mh^2R^2\frac{1}{r^5}\,,
\label{circforcebis}
\eeq
consistently with \eqref{-5force}.

\goodbreak
\section{Outlook and some history}

The correspondence between a harmonic oscillator  and
the Kepler problem, suggested by Arnold \cite{Arnold} following Levi-Civita \cite{LC} and Bohlin \cite{Bohlin} allowed us to deduce the Kepler laws of planetary motion from those, simpler, of an oscillator.
The correspondence can be extended to
scattered motions exemplified by non-periodic comets and  Rutherford scattering. 

A key element  is the redefinition  (\ref{newtime}) of the parameter along each trajectory,  suggested by the comparison  (\ref{bolyoscter}) of the respective areal velocities.

Kepler, in his \textit{``Astronomia Nova''}  \cite{Kepler} deduced his First Law from Tycho's observations of Mars (and then boldly extended to all planets).
Modern textbooks derive in turn the laws of planetary motion from Newton's inverse-square law  \cite{Goldstein}. The question can however be raised also conversely: \textit{What form of a central force can be  consistent with observed trajectories ?}

As an example, one can wonder with what force law would Galilei's (naive) picture in his ``Dialogo'', Fig.\ref{DialogoFig}, consistent~? A simple calculation shows that such a trajectory requires a force proportional to the \emph{minus fifth power} of the distance, (\ref{-5force}). The more general question~: \textit{``Which central force law does allow circular motion~?''} was \emph{answered} by Newton \cite{Principia}, vol I. Proposition VII. Problem II. using elementary geometry. 

At the end of the 17th century, the consistency of the inverse-square law  with circular motion around the Earth' center was known widely ; Newton mentions Hooke, Wren, Huygens. However the consistency of Kepler's \emph{elliptic} trajectories with the inverse-square laww was proved first by Newton \cite{Principia}.
More generally, he  proved that the Law of Universal Gravitation allows for conic sections. His contemporaries may have been wondering if these mathematical possibilities do actually exist in nature. The answer was given by Newton himself: collecting the  observational data  from  all around the Earth, he proved that the Great Comet of 1680 followed a \emph{parabolic} trajectory  \cite{Principia} \footnote{According to present knowledge \cite{1680comet}, the 1680 comet follows an elliptic trajectory with period $> 10000$ years.}.
Astronomers have been discovering weakly hyperbolic comets since the mid-1800s.


Arnold argues that both exemples mentioned above are just particular cases of \emph{dual potentials} \cite{Arnold}, Thm 3 p. 97~: let us assume that we have a central force whose strength is proportional to the distance  raised to the power $a$.
Then the mapping
$w \to z = w^{\alpha}$ carries its trajectories to those in a central field whose strength is proportional to the distance raised to the power $A$, where
\beq
(a+3)(A+3)=4,
\qquad
\alpha = \frac{a+3}{2}\,.
\label{dualstrenth}
\eeq
The potentials $r^a$ and and $r^A$ are called dual.
For example, the harmonic and the Newtonian forces are dual to each other~: they correspond to $a=1$ and $A=-2$
respectively ; then $\alpha =2$ as above.
The $r^{-5}$ potential is \emph{self-dual}~: it corresponds to $a=-5=A$. $\alpha=-1$ i.e. $z=1/w$ is the inversion.

We just mention, in conclusion, that the duality relation
mentioned here can be generalized using the theory of conformal mappings \cite{Arnold,Grant,Nersessian,Grandati,Kothawala,Vyska,Chen}, which can be used to large variety of problems which include, apart of planetary motion, also quantum properties and even general relativity. However their study goes beyond our scope here.

\goodbreak

\end{document}